# Mercury Craters Named after Tajik-Persian Poets: Planetary Nomenclature as a Form of Preserving Cultural Heritage

**Rizoi Bakhromzod**

Institute of Astrophysics, National Academy of Sciences of Tajikistan, Dushanbe, Tajikistan
S. U. Umarov Physical-Technical Institute, National Academy of Sciences of Tajikistan, Dushanbe, Tajikistan
E-mail: rizo@physics.msu.ru

**Abstract.** This paper presents, for the first time in Tajik scientific literature, a systematic account of Solar System toponymic objects named after Tajik-Persian poets, scholars, and cultural figures. We show that nine impact craters on the surface of Mercury — the planet closest to the Sun — have been officially approved by the International Astronomical Union (IAU) in honour of outstanding poets of the Persian-Tajik literary tradition: Rudaki, Saadi, Nizami, Rumi, Navoi, Firdousi, Hafiz, Sanai, and Mahsati. For each crater we provide IAU-approved coordinates, diameters, quadrant designations, dates of approval, and geological characteristics derived from the Mariner 10 and MESSENGER missions. The results are placed in a broader context: the Introduction systematises all known Solar System objects — lunar craters, asteroids, and Enceladus surface features — associated with Tajik-Persian civilisation. The chronological span of crater approvals (1976–2025) mirrors the successive stages of Mercury mapping, from Mariner 10 (~40–45% coverage) to MESSENGER (100% coverage). Geological diversity among the nine craters is striking, ranging from the ancient ~490-km Sanai basin (~3.8–3.9 Ga) to the geologically young Firdousi crater with bright secondary ejecta haloes, and encompassing explosive volcanism (Rumi, Navoi), hollow-bearing terrain (Hafiz), polygonal impact morphology (Nizami), and a high-latitude shadowed floor (Saadi). These findings demonstrate that planetary nomenclature constitutes a durable, politically neutral mechanism for the long-term international recognition of cultural heritage.

## 1. Introduction

A well-established tradition in astronomical nomenclature holds that cosmic objects are named after distinguished scientists, poets, thinkers, composers, and other figures whose intellectual and cultural legacy has exerted a lasting influence on world civilisation [1, 2]. This practice fulfils not only a commemorative function but also a historiographic one, inscribing the contributions of individuals and cultural traditions into the symbolic space of contemporary science.

One of the most tangible manifestations of the global recognition accorded to Tajik-Persian civilisation is the assignment of names from its outstanding representatives to objects throughout the Solar System. As of 2025, this corpus of objects encompasses a star and exoplanets, lunar craters, main-belt asteroids, craters on Mercury, features on the surface of Venus, and toponymic objects on Enceladus, a moon of Saturn.

On the lunar surface, eight impact craters have been officially dedicated to major scholars of the Persian-Tajik world. They are surveyed below in chronological order by the lifetimes of the scientists they honour.

The crater *Al-Khwarizmi* (7.1°N, 107.0°E; diameter 56 km) lies on the lunar far side and is assigned a Nectarian age (~3.9–4.0 Ga) [3]. Its western inner wall is considerably wider than the eastern one; a small central peak at the floor mid-point grades into a low ridge trending north-east. The name was approved by the IAU in 1976 in honour of Muḥammad ibn Mūsā al-Khwārizmī (c. 780 – c. 850), the Tajik mathematician and astronomer from Khwarezm whose works established algebra as an independent science and whose Latinised name gave rise to the word 'algorithm' [1]. Notably, before receiving its official IAU designation, the crater was informally referred to as 'Arabia' during the Apollo 17 mission in 1973 [4].

The crater *Alfraganus* (5.4°S, 19.0°E; diameter 20 km, depth 2.8 km) sits on the lunar near side in the rugged highlands south-west of Mare Tranquillitatis. It has a sharp, well-preserved rim and a floor whose area is roughly half that enclosed by the wall [4, 5]. It was named after Aḥmad al-Farghānī (*Alfraganus*, c. 805–870), a Tajik astronomer from Fergana (present-day Uzbekistan). His principal work, *Kitāb fī jawāmiʿ ʿilm al-nujūm* ('Elements of Astronomy'), translated into Latin in the 12th century, was so authoritative that Christopher Columbus relied on its estimate of Earth's circumference when planning his Atlantic voyage [6, 7].

The crater *Azophi* (22.1°S, 12.7°E; diameter 47 km, depth 3.7 km) occupies the rugged south-central highlands on the near side. Its wide outer rim has a characteristic polygonal outline with rounded corners and a relatively sharp, minimally eroded crest; no central peak is present. It was named after ʿAbd al-Raḥmān al-Ṣūfī (903–986), a Tajik astronomer from Rayy (near modern Tehran) known in Europe as Azophi. His fundamental treatise *Kitāb ṣuwar al-kawākib* ('Book of Fixed Stars', 964) contains the earliest documented observation of the Andromeda Galaxy — described as a 'little cloud' — and the first Arabic mention of the Large Magellanic Cloud. An asteroid, (12621) Alsufi, has also been named in his honour.

The crater *Abul Wáfa* (1.0°N, 116.6°E; diameter 55 km) is located on the far side and was approved by the IAU in 1976. It honours Abū al-Wafāʾ al-Būzjānī (940–998), a mathematician and astronomer born in Būzjān (present-day Iran). Al-Būzjānī laid the foundations of spherical trigonometry, introduced the tangent and secant functions, derived numerous trigonometric identities, and pioneered angular-distance measurement methods of direct astronomical and geodetic significance [9].

The crater *Al-Biruni* (17.9°N, 92.5°E; diameter 77 km) lies near the eastern limb of the near side and was approved by the IAU in 1970. It is named after the polymath Abū Rayḥān al-Bīrūnī (973–1048), born in Khwarezm. Al-Bīrūnī devised a method for determining Earth's radius by measuring the dip angle of the horizon from a mountain summit, calculated the densities of more than 18 minerals, offered early descriptions of phenomena consistent with Earth's rotation about the Sun, and founded the comparative study of Indian and Islamic science. The asteroid (9936) Al-Biruni, discovered in 1986, has also been named in his honour [8, 9].

The crater *Avicenna* (39.7°N, 97.2°W; diameter 74 km) lies on the far side and was approved by the IAU in 1970. It commemorates Abū ʿAlī Ibn Sīnā (980–1037), a Tajik philosopher, physician, and scientist born at Afshana near Bukhara. His *Canon of Medicine* (*Kitāb al-Qānūn fī al-ṭibb*) remained the principal medical textbook in Europe and the East until the 17th century [10].

The crater *Omar Khayyam* (58.0°N, 102.1°W; diameter 70 km) is on the far side, approved by the IAU in 1970. It is named after ʿUmar Khayyām (c. 1048 – after 1122), a Tajik mathematician, astronomer, and poet from Nishapur. Khayyām devised a geometric method for solving cubic equations via conic-section intersections and reformed the Iranian solar calendar (1079), achieving an accuracy of 365.24219858 days — more precise than the Gregorian calendar. The MPC citation describes him as a '*great Tadjik and Persian poet, mathematician and philosopher*' — an official international acknowledgement of his dual cultural identity [8]. The asteroid (3095) Omarkhayyam was named in his honour [9].

The crater *Nasireddin* (41.0°S, 0.2°E; diameter 52 km, depth 3.0 km) is on the near side in the rugged southern highlands and is among the oldest in this list by approval date: 1935. The crater overlaps earlier formations — Miller to the north and Huggins to the west. It retains well-preserved morphological detail, including terraced inner walls, a sharp southern and eastern rim, and a roughly level but rough-surfaced floor with several low central peaks.

It was named after Naṣīr al-Dīn al-Ṭūsī (1201–1274), the Persian polymath who founded the Maragha Observatory (1259), developed the mathematical device known as the 'Ṭūsī couple' (which circumvented the equant in Ptolemaic theory and anticipated Copernican kinematics), and compiled the planetary tables known as the *Zīj-i Īlkhānī* [8].

Turning to asteroids, among those named after outstanding representatives of the Persian-Tajik scientific tradition the most prominent is the main-belt asteroid (3095) *Omarkhayyam* (provisional designation 1980 RT$_2$), discovered on 8 September 1980 by L. V. Zhuravleva at the Crimean Astrophysical Observatory and named after Omar Khayyam (citation published 31 May 1988) [9, 11]. The asteroid (2755) *Avicenna* (1973 SJ$_4$) was discovered on 26 September 1973 by L. I. Chernykh at Nauchny (Crimea) and named on 28 March 1983 [10]. Asteroid (9936) *Al-Biruni* was found on 8 August 1986 by Eric Elst and Violeta Ivanova at Rozhen Observatory (Bulgaria); its name citation was published on 26 September 2007 [9]. Asteroid (90806) *Rudaki* (1995 AE), discovered on 4 January 1995 by V. S. Casulli and named on 5 October 2017, honours Rudaki (858–941) as the progenitor of classical Tajik-Persian poetry [9]. Asteroid (10269) *Tusi* was discovered on 24 September 1979 at the Crimean Astrophysical Observatory by N. S. Chernykh and named on 9 March 2001 [9]. Asteroid (11156) *Al-Khwarizmi* was discovered on 31 December 1997 in Prescott, Arizona, continuing the same symbolic recognition of the Persian-Islamic scientific heritage embodied by Avicenna, al-Biruni, and Khayyam [9]. Asteroid (492786) *Ferdowsi*, discovered on 24 September 2009 at Zelenchukskaya by T. V. Kryachko and B. Satovsky and formally cited in the WGSBN Bulletin, honours Firdowsi [9]. The asteroid (12621) *Alsufi* commemorates al-Ṣūfī [9]. Asteroid (12610) *Hāfez*, discovered on 24 September 1960 during the Palomar-Leiden survey, is named after Hafez of Shiraz [9].

A distinct group of asteroids honours Tajikistan itself and its astronomical institutions. The asteroid (2469) *Tajikistan* (1970 HT), discovered on 27 April 1970 by T. M. Smirnova, was named directly after the country [12]. The asteroid (2746) *Hissao* (1979 XX), discovered on 22 September 1979 by N. S. Chernykh and named on 28 March 1983, honours the Hissar (Gissar) Astronomical Observatory, a station of the Institute of Astrophysics of Tajikistan established in 1956 [9, 13]. Among asteroids dedicated to astronomers connected with the Institute of Astrophysics in Dushanbe: (3013) *Dobrovoleva* (1979 SD$_7$, discovered 23 September 1979, named 18 September 1986) honours Oleg V. Dobrovol'sky; (3945) *Gerasimenko* (discovered 14 August 1982 by N. S. Chernykh, named 18 February 1992) honours Svetlana I. Gerasimenko — Dushanbe-based comet researcher and co-discoverer of 67P/Churyumov-Gerasimenko [9, 13]; (4208) *Kiselev* (1986 RQ$_2$, named 11 March 1990) honours Nikolai N. Kiselev, head of a department at the Astrophysical Institute in Dushanbe; (4207) *Chernova* (1986 RO$_2$, named by E. Bowell) honours Galina P. Chernova, a researcher at the same institute [9, 13]; (4011) *Bakharev* (1978 SC$_6$, discovered 28 September 1978) honours Anatoly Bakharev of the Institute of Astrophysics of Tajikistan [9, 13]; (7164) *Babadzhanov* (1984 ET, discovered 6 March 1984) honours Pulod Babadzhanov — director of the Institute of Astrophysics and specialist in meteors and meteor streams [13–15]; (3436) *Ibadinov* (1976 SS$_3$, discovered 24 September 1976) honours Khursandkul Ibadinov, founder of the laboratory for simulation of cometary processes in Dushanbe [13, 16, 17]; and (24533) *Kokhirova* (2001 CR$_{27}$, discovered 2 February 2001 by the Lowell Observatory Near-Earth-Object Search programme, named 2014) honours Gulchehra Kokhirova [13, 18].

In conclusion, Tajik-Persian civilisation is represented in minor-planet nomenclature simultaneously at several levels: through the names of classical scholars and poets — Omar Khayyam, Ibn Sina, al-Biruni, Rudaki, Tusi, al-Khwarizmi, Firdowsi, and Hafez — and through names connected with Tajikistan itself, its observatories, and the astronomers who worked in Dushanbe. This aggregation of asteroid names constitutes a distinctive memorial stratum within astronomical nomenclature, reflecting recognition of the Tajik-Persian world's contribution to science and culture.

A rich layer of Persian-Tajik cultural toponymy is present on the surface of Enceladus, a moon of Saturn. Following the decision of the IAU Working Group for Planetary System Nomenclature (WGPSN), all surface features on Enceladus are named after characters and places from the Arabic-Persian literary monument *One Thousand and One Nights* — a work rooted deeply in the Persian-Tajik narrative tradition [19]. Approved features include the sulci *Baghdad Sulci*, *Samarkand Sulci*, and the craters *Ali Baba*, *Sindbad*, *Aladdin*, and others. Scientists documenting the dense network of Enceladus sulci associated with water-ice jets ('tiger stripes') routinely employ these official IAU toponyms within the framework of the Cassini mission [20].

The naming contribution of Tajik-Persian civilisation to Solar System planetary nomenclature is thus exceptionally broad. Its most systematic representation is found on the surface of Mercury — the planet nearest the Sun — where nine craters bear the names of Persian-language poets. Those objects are the principal subject of this paper.

## 2. Mercury: Physical Characteristics and History of Exploration

Mercury is the smallest planet in the Solar System (diameter 4879.4 km, mass $3.302 \times 10^{23}$ kg) and the closest to the Sun (mean distance $57.91 \times 10^6$ km, or 0.387 AU) [21]. In the Iranian Islamic astronomical tradition the planet was known as *Tīr (تیر)* — derived from Zoroastrian astral lore — and under the Arabic name *ʿUṭārid (عطارد)*. Mercury completes one orbit in 87.97 Earth days, locked in a unique 3:2 spin-orbit resonance (sidereal rotation period 58.646 days), so that a solar day on Mercury lasts approximately 176 Earth days [21]. Mercury's orbital eccentricity (0.2056) is the largest among the planets, making it the critical test case for Ptolemaic and later cosmological models. Its surface temperature range — from −180°C (night) to +430°C (day) — is the most extreme in the Solar System [21].

Three space missions have directly explored Mercury's surface. NASA's *Mariner 10* (launched 3 November 1973) performed three planetary flybys in 1974–1975, mapping roughly 40–45% of the surface [22] and providing the data underpinning the initial planetary nomenclature. The orbital spacecraft *MESSENGER* (NASA; launched 3 August 2004; in Mercury orbit 18 March 2011 – 30 April 2015) acquired approximately 300,000 images with 100% surface coverage, detecting water ice at the poles, unique hollow-forming terrain, evidence of extensive volcanism, and global tectonic contraction [23]. The European-Japanese mission *BepiColombo* (ESA/JAXA; launched 20 October 2018) completed six Mercury gravity assists; orbital insertion is expected in November 2026 [24].

## 3. The IAU Naming System for Mercury Impact Craters

The nomenclature for Mercury's surface features was developed by the IAU Working Group for Planetary System Nomenclature (WGPSN). Although designations after birds or towns were initially considered, a different concept was adopted in 1975, driven in large part by the active advocacy of the American astronomer Carl Sagan [25]. The resolution recorded in the proceedings of the XVI IAU General Assembly (Grenoble, 1976) reads: *'Resolved that the large craters on Mercury be named after great contributors to the humanities and the arts; including (but not limited to): authors of drama, prose and poetry; painters; sculptors; architects; composers; and musicians'* [26]. The canonical paper codifying this system in the scientific literature is Morrison (1976) [27].

Naming rules stipulate that the honoured individual must have been deceased for at least three years at the time of submission; their renown must span at least fifty years; names of political or military figures and religious leaders are excluded. The requirement for international representation across ethnic groups has resulted in extensive inclusion of Persian-Tajik, Arabic, Indian, Chinese, and other non-Western cultural figures [26, 27].

Mercury's surface is divided into fifteen quadrants (H-1 to H-15), each mapped at 1:5,000,000 scale (Fig. 1). Longitudes are measured from the crater Hunt/Hunkal as the prime meridian origin in the modern IAU system. As of 2025, a total of 444 craters have been officially approved [8], nine of which bear the names of Persian-Tajik poets.

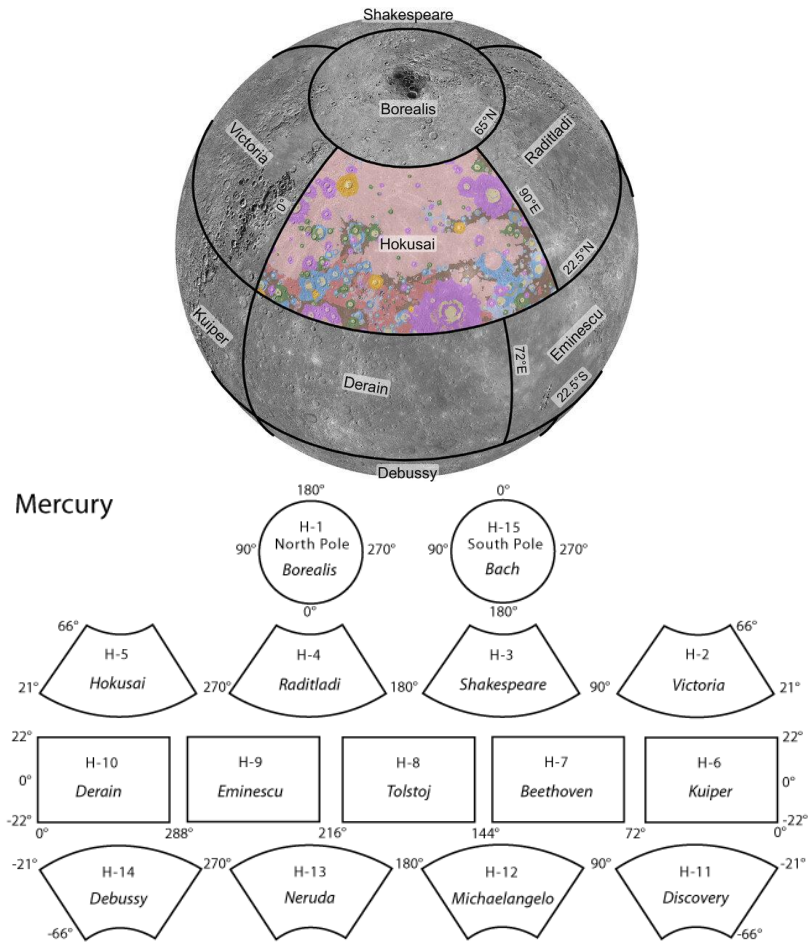

Fig. 1. (A) Schematic division of Mercury's surface into cartographic quadrants by latitude and longitude; (B) layout of the fifteen Mercury quadrant maps (H-1 to H-15). Source: IAU WGPSN / USGS Astrogeology Research Program.

## 4. Mercury Craters Bearing the Names of Persian-Tajik Poets

### *4.1  Rūdakī (ID 5252)*

Crater Rūdakī lies in the equatorial zone of Mercury (3.97°S, 51.76°W; diameter 124 km; Kuiper quadrant, H-6) and is among the first surface features to receive an official designation, in 1976 [8]. The crater floor is mantled by the smooth volcanic plains of *Sihtu Planitia* — the largest lava plain in the region (~565 km in extent) — indicating post-impact basaltic flooding that buried the original floor [28]. During the BepiColombo flyby of 1 October 2021 this area was imaged as a priority target (Fig. 2) [24].

The crater is named after **Abū ʿAbdullāh Rūdakī** (c. 858–941), the founder of Classical Tajik-Persian literature, born in Panjrud near Penjikent (present-day Tajikistan). Court poet of the Samanid Emir Nasr II in Bukhara, Rūdakī is credited with composing more than 180,000 couplets, of which only a few hundred survive. The French Iranist F. de Blois described him as 'the most celebrated Persian poet before Firdowsi'.

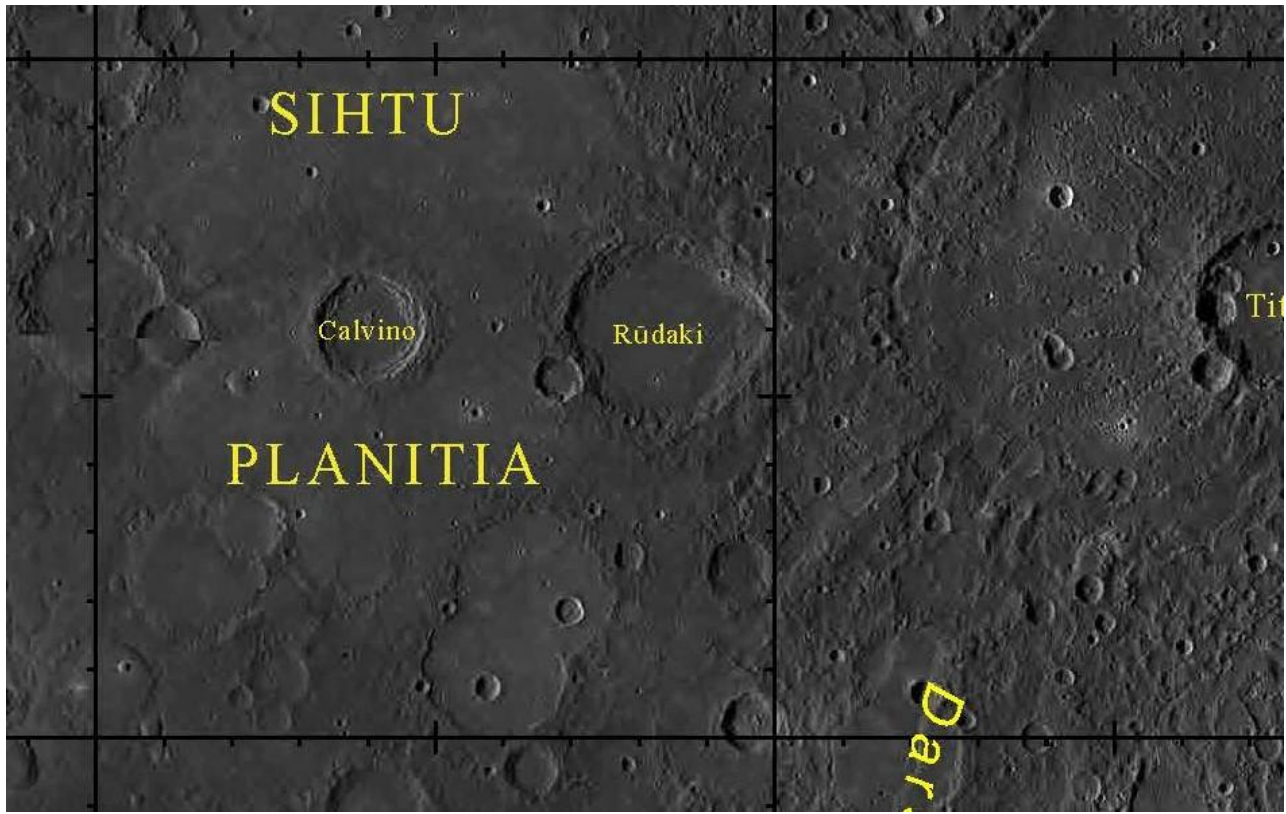

Fig. 2. Crater Rūdakī on the surface of Mercury as shown in the IAU Gazetteer of Planetary Nomenclature (USGS/IAU, 2024). The crater floor is almost entirely buried by lavas of Sihtu Planitia.

### *4.2 Sadī (ID 5271)*

Crater Sadī (79.24°S, 51.26°W; diameter 66.5 km) was approved by the IAU in 1976 and lies in the polar quadrant Bach (H-15) [8]. Its high southern latitude implies the potential presence of permanently shadowed regions on its floor — analogous to those in which water ice has been detected on Mercury [23]. The crater is named after **Muslihuddin Saadi Shirazi** (c. 1210–1292), author of *Gulistān* (1258) and *Bostān* (1257). Saadi's verse 'All human beings are members of one body, for all are created from the same essence' is woven into a carpet in the United Nations headquarters in New York.

*4.3 Nizāmī (ID 4315)*

Crater Nizāmī (70.46°N, 166.71°W; diameter 77 km; Borealis quadrant, H-1) was approved by the IAU in 1979 [8]. Morphologically it is classified as a *polygonal impact crater (PIC)*: its wall contains rectilinear segments reflecting pre-existing fault systems in the lithosphere. It commemorates **Nizami Ganjavi** (c. 1141–1209), born in Ganja, author of the *Khamsa* ('Quintet', ~30,000 couplets).

*4.4 Rūmī (ID 5253)*

Crater Rūmī (24.2°S, 105.3°W; diameter 75 km; Michelangelo quadrant, H-12) was approved by the IAU in 1985 [8]. Geologically it is particularly noteworthy: an irregular pit at the crater centre, surrounded by brighter pyroclastic material, provides evidence of **explosive volcanism** [11]. The thrust-fault scarp Palmer Rupes, associated with Mercury's global tectonic contraction, crosses the crater [12]. Thomas et al. (2014) catalogued the volcanic vent inside Rūmī as one of ~188 sites of long-lived explosive volcanism on the planet [11]. The crater honours **Jalāluddīn Rūmī** (1207–1273), author of the *Masnavi* (~25,000 couplets).

*4.5 Navoi (ID 14514)*

Crater Navoi (58.82°N, 160.41°E; diameter 69 km; Raditladi quadrant) was approved by the IAU in November 2008 on the basis of data from the first MESSENGER flybys [8]. It is a classical *pit-floor crater*: an irregular central depression and anomalous reddish material of unusual mineralogy are interpreted as evidence of shallow magmatic activity and probable roof collapse above a drained magma reservoir [13]. The crater is named after **Alisher Navoi** (1441–1501), the founder of Uzbek literature and author of the *Khamsa* in Chagatai Turkic — the first complete 'quintet' in any Turkic language. Navoi also composed works in Persian.

*4.6 Firdousi (ID 14648)*

Crater Firdousi (3.47°N, 65.32°E; diameter 98 km; Derain quadrant, H-10) was approved by the IAU on 3 March 2010 in a package of ten features named on the basis of the third MESSENGER flyby (29 September 2009) [8]. Its flat floor indicates **volcanic lava flooding** following impact, which has almost entirely buried the central peak (Fig. 3) [28]. High-reflectance bluish haloes of secondary crater chains imply a relatively young age.

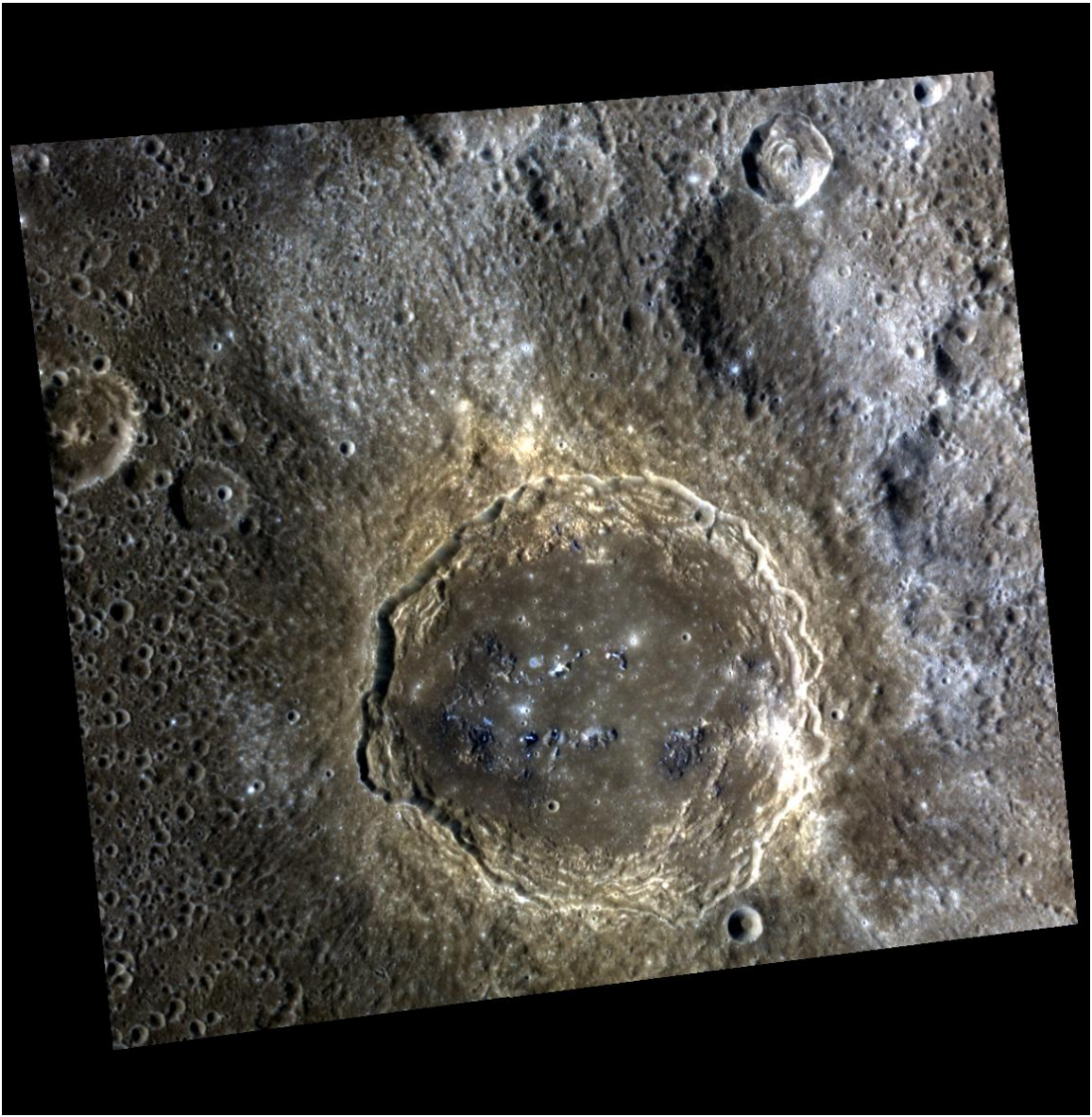

Fig. 3. Crater Firdousi on the surface of Mercury. Image acquired on 14 August 2011 by the Wide Angle Camera (WAC) of the Mercury Dual Imaging System (MDIS) aboard MESSENGER; filters at 996, 748, and 433 nm (RGB composite); spatial resolution 204 m pixel$^{-1}$; crater diameter ~96 km.

In the IAU Gazetteer the crater is described as honouring '*Abulkasim Firdousi, Tajik/Persian poet (c. 940–1020/30)*' — a formulation that fixes the Tajik-Persian identity of the poet at the level of an international astronomical standard. **Firdowsi** (Abolqāsem Manṣūr, c. 940–1020/1025) is the author of the *Shāhnāme* ('Book of Kings', ~50,000 couplets, completed 8 March 1010), the greatest epic in the Persian language. UNESCO marked the millennium of the *Shāhnāme* in 2010–2011.

### *4.7 Hafiz (ID 15245)*

Crater Hafiz (19.5°N, 79.56°E; diameter 280 km) was approved by the IAU on 23 June 2014 [8]. It is one of the two largest 'Persian' objects on Mercury. A smaller interior crater contains a dark low-reflectance material (LRM) patch associated with **hollows** — features unique to Mercury formed by sublimation of volatile compounds (likely sulphides) under solar radiation [23]. Blewett et al. (2011) described hollows as evidence of geologically recent volatile-loss activity [23]. The crater commemorates **Hafez of Shiraz** (c. 1315–1390), the unsurpassed master of the ghazal and author of the *Divan* (~500 ghazals). Goethe, inspired by his verse, composed the *West-Eastern Divan* (1819).

## *4.8 Sanai (ID 15246)*

Crater Sanai (13.37°S, 6.99°W; **diameter 490 km**; Kuiper quadrant, H-6) was approved by the IAU in 2014 [8] and is the largest feature of the Persian-Tajik group on Mercury. It is an *ancient impact basin of Tolstojan age* (~3.8–3.9 Ga), characterised by radial lineations to the north-east (analogous to the 'Imbrian sculpture' on the Moon) and a floor blanketed by smooth lava plains [28]. Prior to its naming the basin was catalogued as 'b38'. It honours **Ḥakīm Sanāʾī of Ghazni** (c. 1080–1131), author of the *Ḥadīqat al-ḥaqīqa* ('Garden of Truth', ~10,000 verses) — the first major Sufi masnavi. Rumi wrote of him: 'Attar was the spirit, Sanai his eyes, and we came in their wake'.

## *4.9 Mahsati (ID 16387)*

Crater Mahsati was approved by the IAU on 7 February 2025 as part of a package of seven new Mercury features [8] and is the most recent addition to the 'Persian collection' of planetary nomenclature. Its precise coordinates and diameter are published in the IAU Gazetteer (Feature 16387). The crater is named after **Mahsati Ganjavi** (c. 1089–1159), one of the most significant women poets of medieval Persian literature, master of the *rubāʿī* (quatrain), and court poet of the Seljuk sultan Ahmad Sanjar. UNESCO marked the 900th anniversary of her creative legacy in 2013.

The Persian-Tajik component of Mercury's nomenclature extends beyond the nine craters honouring poets. Also significant is *Tir Planitia* (ID 6025, approved 1976; 1.04°S, 176.69°W; longest dimension ~754 km), where the IAU Gazetteer explicitly states that the name *Tir* is the Persian word for the planet Mercury — thus inscribing the Persian name of the planet itself into its own surface nomenclature. Additionally, the crater *Ustad Isa* (ID 6249, approved 1979; 31.91°S, 166.11°W; diameter 138 km) is attributed in the official nomenclature to a 17th-century Persian-Turkish architect traditionally associated with the builders of the Taj Mahal, though his precise design role remains debated.

**Table 1**

*Mercury features approved by the IAU associated with Persian-Tajik cultural heritage [8, 27]*

| Crater | Poet (dates) | Centre coords (lat., lon.) | Diam. (km) | Quadrant | IAU year | Geological notes |
|---|---|---|---|---|---|---|
| Rūdakī (ID 5252) | Abū ʿAbdullāh Rūdakī (c. 858–941) | 3.97°S 51.76°W | 124 | Kuiper (H-6) | 1976 | Smooth volcanic plains of Sihtu Planitia; lava-flooded floor |
| Sadī (ID 5271) | Muslihuddin Saadi Shirazi (c. 1210–1292) | 79.24°S 51.26°W | 66.5 | Bach (H-15) | 1976 | Near south pole; perma-shadow on floor; possible water-ice reservoir |
| Nizāmī (ID 4315) | Nizami Ganjavi (c. 1141–1209) | 70.46°N 166.71°W | 77 | Borealis (H-1) | 1979 | Polygonal impact crater (PIC); rectilinear wall segments indicate pre-existing fault fabric |
| Rūmī (ID 5253) | Jalāluddīn Rūmī (1207–1273) | 24.2°S 105.3°W | 75 | Michelangelo (H-12) | 1985 | Explosive vent / pyroclastic deposit at centre; Palmer Rupes thrust fault crosses the crater |
| Navoi (ID 14514) | Alisher Navoi (1441–1501) | 58.82°N 160.41°E | 69 | Raditladi | 2008 | Pit-floor crater; anomalous reddish material; probable magmatic collapse pit |
| Firdousi (ID 14648) | Abolqāsem Firdowsi (c. 940–1020) | 3.47°N 65.32°E | 98 | Derain (H-10) | 2010 | Lava-flooded floor; buried central peak; bright blue secondary-crater chains |
| Hafiz (ID 15245) | Hafez of Shiraz (c. 1315–1390) | 19.5°N 79.56°E | 280 | Eminescu | 2014 | Large impact basin; low-reflectance material (LRM); hollows indicating recent volatile loss |
| Sanai (ID 15246) | Ḥakīm Sanāʾī of Ghazni (c. 1080–1131) | 13.37°S 6.99°W | 490 | Kuiper (H-6) | 2014 | Ancient impact basin (~3.8–3.9 Ga); radial lineations; smooth volcanic floor; formerly 'b38' |
| Mahsati (ID 16387) | Mahsati Ganjavi (c. 1089–1159) | 40.50°N 249.18°E | 79 | H-04 | 2025 | Approved 7 Feb 2025; detailed characterisation pending BepiColombo orbital data |
| Tir Planitia (ID 6025) | Tir — Persian name for Mercury | 1.04°S 176.69°W | 754* | H-08 | 1976 | Volcanic plain (planitia); *longest dimension |
| Ustad Isa (ID 6249) | Ustad Isa — Persian-Turkic architect (17th c.) | 31.91°S 166.11°W | 138 | H-12 | 1979 | Impact crater |

## 5. Discussion

The dataset assembled here permits several observations. First, the nine Mercury craters bearing Persian-Tajik poets' names span an IAU-approval interval of nearly fifty years (1976–2025), covering virtually the entire history of Mercury nomenclature. The chronological clustering — two craters in 1976, one in 1979, one in 1985, one in 2008, one in 2010, two in 2014, one in 2025 — directly mirrors the staged expansion of Mercury cartographic coverage: Mariner 10 (1974–1975) mapped ~40–45% of the surface, while MESSENGER (2011–2015) achieved complete global coverage and generated a data torrent that drove subsequent nomenclature expansion.

Second, the geological diversity of the craters is as rich as the literary heritage of their eponyms. The giant Sanai basin (490 km) ranks among the largest tectono-volcanic structures on the planet and exhibits characteristics consistent with formation during the early epoch of Mercury's history [28]. In contrast, the comparatively young Firdousi crater, with its bright blue secondary ejecta haloes, testifies to relatively late volcanic activity [23]. The Rumi crater, with its pyroclastic vent and the Palmer Rupes scarp, is one of the best-studied sites of long-lived explosive volcanism [11, 12]. The Hafiz crater with hollows serves as a natural laboratory for studying geologically active volatile-loss processes [23].

Third, it is noteworthy that the IAU Gazetteer formally describes the Firdousi crater as dedicated to a 'Tajik/Persian poet' — a formulation that acknowledges dual cultural identity and underscores the inseparability of Tajik heritage from the broader Persian literary tradition. An analogous characterisation applies to Rūdakī, born on the territory of present-day Tajikistan.

The broad scope of Solar System toponymy linked to Persian-Tajik civilisation — the lunar craters al-Biruni, Avicenna, Omar Khayyam, Abul Wáfa; the asteroids (3095) Omarkhayyam, (2755) Avicenna, (9936) Al-Biruni, (2746) Hissao, (2469) Tajikistan, and those honouring Babadzhanov, Ibadinov, and Kokhirova; the Enceladus naming scheme derived from *One Thousand and One Nights* — collectively testify to the comprehensive recognition by the international scientific community of this civilisation's role in the history of science, culture, and literature. Planetary nomenclature constitutes a unique mechanism for the durable, politically neutral international inscription of cultural heritage.

**6. Conclusions**

This study yields the following conclusions:
1. Nine impact craters on the surface of Mercury have been officially approved by the IAU in honour of Persian-Tajik poets: Rūdakī (124 km, 1976), Sadī (66.5 km, 1976), Nizāmī (77 km, 1979), Rūmī (75 km, 1985), Navoi (69 km, 2008), Firdousi (98 km, 2010), Hafiz (280 km, 2014), Sanai (490 km, 2014), and Mahsati (2025). All are recorded in the IAU Gazetteer of Planetary Nomenclature [8].

2. The current system of naming Mercury craters after figures from the humanities was adopted by the IAU in 1976, with the direct involvement of Carl Sagan, and is documented in Morrison (1976) [27]; its foundational principle is set out in Transactions of IAU XVIB [26].

3. The craters display a wide spectrum of geological characteristics: lava flooding (Rūdakī, Firdousi, Sanai), explosive volcanism (Rūmī, Navoi), hollows (Hafız), polygonal impact morphology (Nizāmī), and near-polar positioning with possible permanent shadow (Sadī).

4. These findings may inform science-education and public outreach initiatives in Tajikistan and may serve as a foundation for developing proposals to name additional Solar System objects after outstanding figures of Tajik culture not yet represented in planetary nomenclature.

**Acknowledgements**

The author thanks the staff of the USGS Astrogeology Science Center for maintaining the publicly accessible IAU Gazetteer of Planetary Nomenclature and the IAU Minor Planet Center for open access to the MPC Orbit Database. BepiColombo is a joint ESA/JAXA mission.